\documentclass[10pt,a4paper, twocolumn]{article}
\pdfoutput=1
\usepackage[utf8]{inputenc}
\usepackage{amsmath}
\usepackage{amsfonts}
\usepackage{amssymb}
\usepackage{graphicx}

\usepackage{amsfonts, amscd, amsmath, amssymb}
\usepackage{url,ulem}
\usepackage{graphicx, epsfig}
\usepackage{multirow}

\usepackage[left=2cm,%
                right=2cm,%
                top=2.25cm,%
                bottom=2.25cm,%
                headheight=12pt]{geometry}%
\usepackage{color}

\usepackage{bm}
\usepackage[vlined,ruled]{algorithm2e}
\usepackage[multiple]{footmisc}

\SetKwInOut{Input}{Input}
\SetKwInOut{Output}{Output}
\SetKwInOut{Parameters}{Parameters}
\SetKwComment{Comment}{}{}

\usepackage{amsmath}
\usepackage{amsfonts}
\usepackage{amssymb}
\usepackage{color}

 \DeclareGraphicsExtensions{.pdf,.jpeg,.png,.jpg}
 \usepackage[multidot]{grffile}
 
\usepackage{subcaption}
\usepackage{array}
\newcolumntype{?}[1]{!{\vrule width #1}}

\usepackage{makecell}
\usepackage{enumitem}

\makeatletter
\newcommand\footnoteref[1]{\protected@xdef\@thefnmark{\ref{#1}}\@footnotemark}
\makeatother
\usepackage{pgf,tikz}
\usepackage{mathrsfs}
\usepackage{pgfplots}
\usetikzlibrary{arrows}

\usepackage{makecell}
\usepackage{enumitem}
\usepackage{changepage}

\usepackage{hyperref}
\title{Efficient joint noise removal and multi exposure fusion}
\author{A. Buades\textsuperscript{1\textpilcrow},
J.L Lisani\textsuperscript{1\textpilcrow},
O. Martorell\textsuperscript{1*\textpilcrow}}
\date{}
\begin{document}
\maketitle
\begin{flushleft}
\bigskip
\textbf{1} Institute of Applied Computing and Community Code (IAC3) and with the Dept.of Mathematics and Computer Science, Universitat de les Illes Balears, Cra.~de Valldemossa km.~7.5, E-07122 Palma, Spain
\\
\bigskip

%
%




\textpilcrow The authors acknowledge the Ministerio de Ciencia, Innovación y Universidades (MCIU), the Agencia Estatal de Investigación (AEI) and the European Regional Development Funds (ERDF) for its support to the project TIN2017-85572-P.

* o.martorell@uib.cat

\end{flushleft}
\section*{Abstract}
Multi-exposure fusion (MEF) is a technique for combining different images of the same scene acquired with different exposure settings into a single image. All the proposed MEF algorithms combine the set of images, somehow choosing from each one the part with better exposure.    

We  propose a novel multi-exposure image fusion chain taking into account noise removal.  
The novel method takes advantage of  DCT processing and the multi-image nature of the MEF problem.  
We propose a joint fusion and denoising strategy taking advantage of spatio-temporal patch selection  and collaborative 3D thresholding.  The overall strategy permits to denoise and fuse the set of images without the need of recovering each denoised exposure image, leading to a very efficient procedure.



\section{Introduction}

Multi-exposure fusion (MEF) is a technique for combining different images of the same scene acquired with different exposure settings into a single image. 
By keeping the best exposed parts of each image,  we can recover a single image where all features are well represented. 
Compared to High Dynamic Range (HDR),  MEF  avoids the estimation of the camera response function and the tone mapping, keeping during all the process the original range of { the} images.

All the proposed MEF algorithms combine the set of images, somehow choosing from each one the part with better exposure.   Most methods express this choice as a combination of pixel values or their gradient, commonly weighted depending on exposure, saturation and contrast, e.g. Mertens et al. \cite{mertens_article}. In order to make this choice robust, the methods use a pyramidal structure or work at the patch level instead of the pixel one.  In the case that pixel gradient is manipulated, a final estimate has to be recovered by Poisson editing \cite{perez2003poisson}. 

Image fusion is an extensive research area not limited to multi-exposure images. The combination of several images permits to improve their quality, removing for example noise  \cite{buades2016patch}, compression artifacts  \cite{haro2012photographing}, haze  \cite{joshi2010seeing}, blur \cite{rav2005two} or shaking blur from hand held video \cite{delbracio15},\cite{delbracio15video}.  

We  propose a novel multi-exposure image fusion chain taking into account noise removal. Instead of combining gradients or pixel values, we fuse the Discrete Cosinus Transform (DCT) coefficients of the differently exposed images. 
The use of the DCT permits to include a thresholding stage attenuating noise during the fusion process. 
DCT thresholding is a well known method for noise removal, behaving extremely well with moderate noise levels and being  efficient in terms of computational complexity.  We propose  a novel strategy taking advantage of DCT thresholding and the multi-image nature of the MEF problem.  

The methods { in \cite{Ma17} and \cite{martorell2019ghosting} propose} a similar fusion procedure to ours. However,  they do not include any noise removal stage.
The method in Ma et al. \cite{Ma17} uses a patch based approach, in which each patch is decomposed into its average and detail patch. Averages are combined depending on exposure{,} while detail patches are averaged depending on its magnitude. Our approach also deals differently with the average and the details but it uses DCT to represent the details and not just the difference of the patch with its average.  While the method in \cite{Ma17} is applied to each color channel, we use a color decorrelation transform. 
 
The method in \cite{martorell2019ghosting}  uses a DCT transform but instead of combining the averages of the patches it uses an auxiliary image obtained with  Mertens et al \cite{mertens_article} algorithm to set the patch illumination. We combine the average coefficients depending on local and global exposure.
The method in  \cite{martorell2019ghosting} applies a completely different strategy for luminance and chromatic parts, while our method uses the same procedure for both components, permitting noise reduction.

Although the proposed fusion method is able to deal with low levels of noise, for more challenging cases the simple DCT thresholding might be insufficient. Based on noise removal methods for image and video   \cite{dabov2007image, buades2016patch}, we propose a joint strategy taking advantage of spatio-temporal patch selection \cite{buades2016patch} and collaborative 3D thresholding  \cite{dabov2007image}.  The computed patch DCT{s} for collaborative thresholding are directly fed into the fusion procedure.  
The overall strategy permits to denoise and fuse the set of images without the need of recovering each denoised exposure image{,}  { leading to a very fast procedure}.


\medskip

Section \ref{sec:related} describes the MEF existing literature. The proposed method for the fusion of { images with} different exposures is described in {S}ection   \ref{sec:proposed}. Section \ref{sec:bm3d} presents the complete method including the fusion and noise removal   { stages}.  {I}n {S}ection \ref{sec:discussion}, we discuss the implementation of the method and compare with state of the art algorithms { and we draw some conclusions in Section~\ref{sec:conclusions}.}

\section{Related work} \label{sec:related}

\graphicspath{{figures_2/sourceImages/}}
\begin{figure*}[t!]
\centering
\includegraphics[width=3cm]{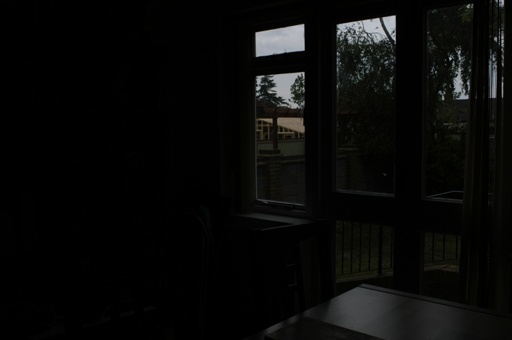} 
\includegraphics[width=3cm]{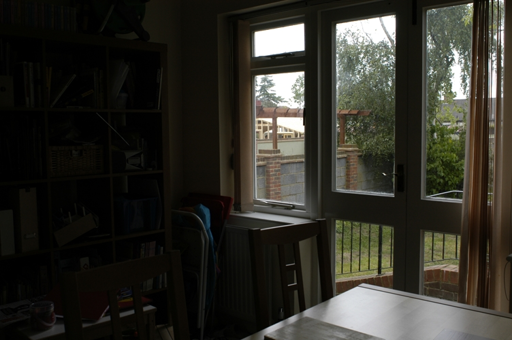} 
\includegraphics[width=3cm]{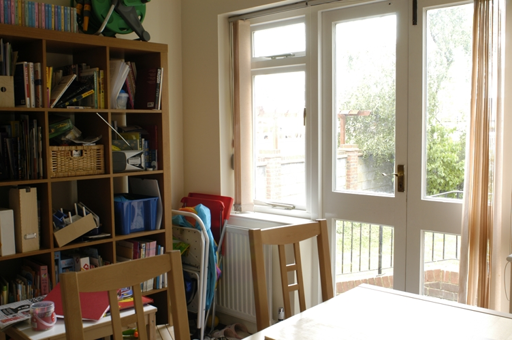} 
\includegraphics[width=3cm]{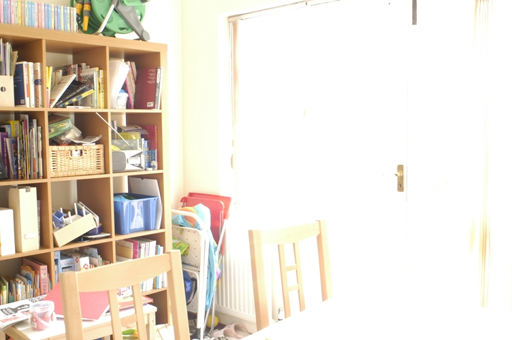} 

\includegraphics[width=3cm]{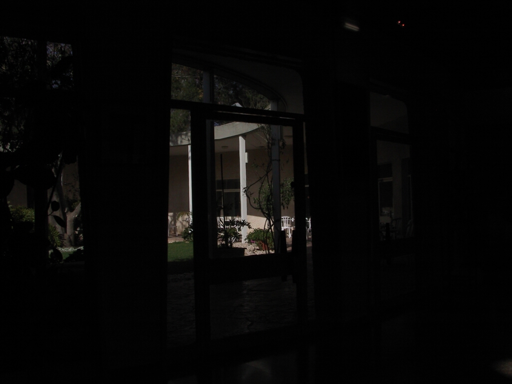} 
\includegraphics[width=3cm]{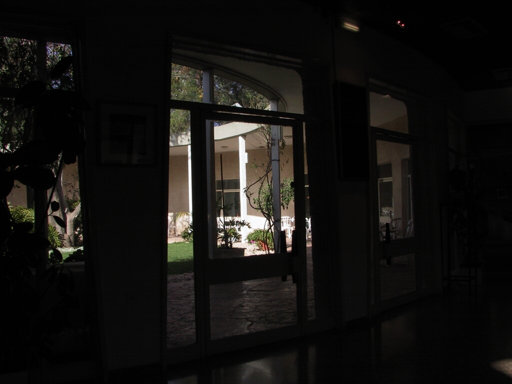} 
\includegraphics[width=3cm]{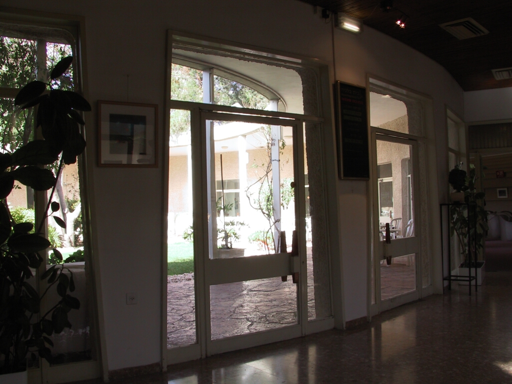} 
\includegraphics[width=3cm]{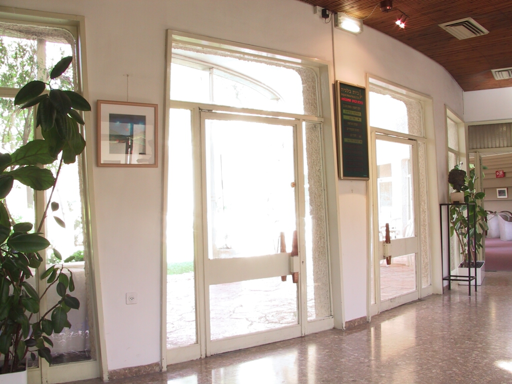} 

\includegraphics[width=3cm]{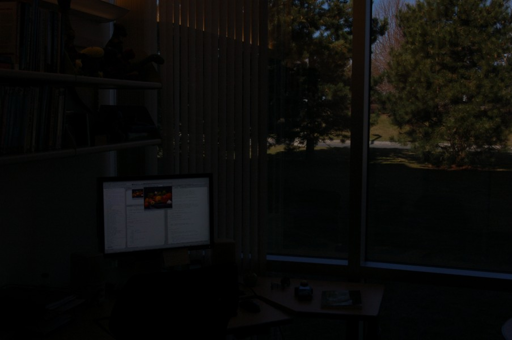} 
\includegraphics[width=3cm]{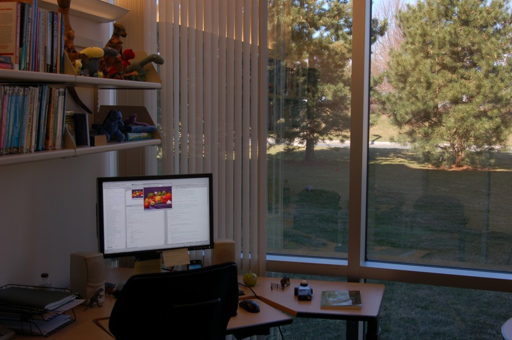} 
\includegraphics[width=3cm]{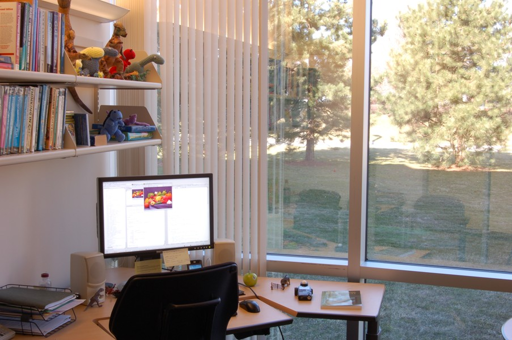} 
\includegraphics[width=3cm]{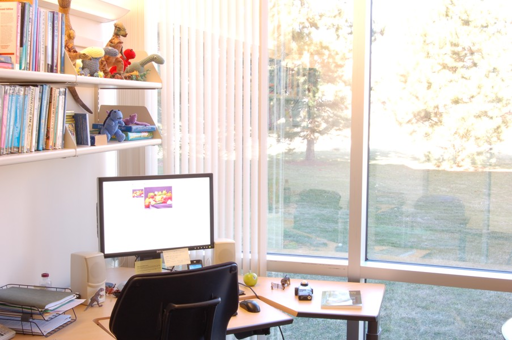} 

\caption{Multi exposure data sets used for comparison.} 
\end{figure*}

\medskip
{ Several MEF methods have been proposed in the literature.}
Mertens et al. \cite{mertens_article} proposed a pixel wise combination { of images} depending on contrast, saturation and well-exposedness. 
Many algorithms appeared improving aspects of their method  or making it motion adaptive, { such as} Li et al. \cite{SLi12}, An et al. \cite{an2011multi}, Ocampo et al. \cite{ocampo2016non}, Liu et al. \cite{liu2015dense}, Hayat et al. \cite{HAYAT2019295} and Hessel et al. \cite{ipol.2019.278}.

Another type of methods combine the gradient of the images, and then they recover the solution by Poisson image editing \cite{perez2003poisson}.  
For several algorithms, just the most convenient gradient is chosen, for example Kuk et al. \cite{kuk2011high}. Other methods combine all gradients,  for example \cite{raskar2005image, zhang2012gradient, Gu12, sun2013poisson}. 

Instead of fusing red, green and blue values, the YCbCr \cite{gonzalezdigital} color space is often used.
Paul et al. \cite{paul2016multi} combines the luminance using a gradient based fusion. 
The chromatic components are  fused by direct pixel-wise averaging fav{o}ring col{o}red pixels compared to gray ones. A similar approach is adopted in \cite{martorell2019ghosting}.

\medskip

This pixel-wise choice is not robust enough and needs to be regularized by using pyramidal structures or working at the patch level. 
The Laplacian pyramid proposed by Burt and Adelson \cite{burt83} was adopted by  Mertens et al. \cite{mertens} and posterior methods Ancuti et al. \cite{ancuti2017single}, Kou et al. \cite{kou2018edge, kou2017multi}.
Patch-based methods use as minimal unit small image windows, which makes them more robust than pixel wise algorithms since much more values are involved  { in the fusion process (}Goshtasby  \cite{goshtasby05}, Zhang et al. \cite{zhang2017patch}, Zhang et al. \cite{zhang2017motion}, Ma et al. \cite{Ma17}{)}.
 
\medskip

When the image sequence is not completely static, the fusion of moving objects creates a ghosting effect. 
There exist three main strategies to avoid these artifacts. The first strategy is to take into account block distance or correlation  in order to weight the combination process.  Such a weighted average  might discard moving objects from the combination. The second one is to modify the sequence to make it static {\cite{hu2013hdr,li2014selectively,zheng2015superpixel,martorell2019ghosting}}.  A third straightforward strategy, adopted by several algorithms \cite{tico2010motion,prabhakar2016ghosting, sie2014alignment, hu2012exposure}, is to fuse several instances of the same reference image. The image with middle exposure is chosen as reference and is photometrically matched to each of the rest. These copies of the reference are then fused by any MEF method.
This strategy permits to get rid of motion and misalignments, but does not use the detail and structure information of the rest of images,  only their color distribution. 

\medskip

 Several methods { (Z. G. Li et al. \cite{Li12, Li17}, Singh et al. \cite{singh14}, Raman et al. \cite{raman09}, Li et al. \cite{SLi13})} separate low frequency (base images) from high frequency (detail images)  using the Bilateral filter \cite{tomasi1998bilateral} or the Guided filter \cite{he2013guided}.  This separation permits an additional enhancement of the detail part.

\medskip

Variational methods \cite{piella2009image, bertalmio2013variational, hafner2015variational} have also been proposed to model MEF fusion. The proposed energies prefer to keep the geometry of the short  exposure images and the color distribution of the long ones.  The methods \cite{ma2015high, laparra2017perceptually,ma2018multi}  try to optimize a quality measure of the final estimate.
Ma et al. \cite{ma2018multi} proposed to optimize a modification of the objective quality measure in \cite{ma2015perceptual}, adapted to color images. 

\medskip

Neural network methods have recently appeared for MEF and HDR imaging. Prabhakar et al.  \cite{prabhakar2017deepfuse} uses a loss function minimizing an exposure fusion objective metric  \cite{ma2015perceptual},  avoiding the creation of ground truth images for the learning phase. Prabhakar et al. \cite{prabhakar2019fast_fusion} proposed a neural network to align differently exposed images and merge them. 
Kalantari et al. \cite{kalantari2017deep} and Wu et al. \cite{wu2018deep} proposed learning methods for HDR imaging.  Both networks apply to sequences of three images, and for the learning phase they need the corresponding LDR and HDR pairs for the three images.  The main difference between both methods is that in \cite{kalantari2017deep} the input images are aligned by an optical flow estimation and warping, whereas in \cite{wu2018deep} the images do not need to be aligned before being introduced to the network.

 Zhang et al. \cite{ZHANG202099} and Xu et al. \cite{xu2020u2fusion} proposed   a unified deep learning framework for several fusion tasks, including MEF.  Zhang et al. \cite{ZHANG202099} extract features using convolutional layers, which are combined using an elementwise operation (mean for multi-exposure images). The final result is reconstructed from the fused features by additional convolutional layers.  
 Xu et al. \cite{xu2020u2fusion} trained a neural network to preserve the adaptive similarity between the fusion result and source exposures, that is, not requiring the ground truth images.
Li et al. (CNNFEAT) \cite{8451689} use CNN to extract features, which are used to combined the different exposures. However, the fused result is not the direct output of a neural network.

There is very few literature dealing with noise removal during multi exposure image fusion, and most
{ published papers} are focused on HDR.
Akyuz et al. \cite{akyuz2007noise} denoise each frame before fusion, but this is performed in the radiance domain. 
Tico et al. \cite{tico2010motion} combines an initial fusion with the { image of the sequence with the shortest exposure} in the luminance domain.
This combination is performed in the wavelet domain and coefficient attenuation is applied to the coefficients of the difference of luminance{s}. Min et al. \cite{min2011noise}  filter the set of images by spatio-temporal motion compensated anisotropic filters prior to HDR reconstruction. Lee et al. \cite{lee2014ghost} use sub-band architecture for fusion, with a weighted combination using a motion indicator function to avoid ghosting effects.  The low frequency bands are filtered with a multi-resolution bilateral filter while the high frequency bands are filtered by soft thresholding. Ahmad et al. \cite{ahmad2016noise}  identify noisy pixels and reduce their weight during image fusion.

\section{Proposed fusion algorithm}\label{sec:proposed}

We propose a novel algorithm for multi-exposure fusion adopting the Fourier aggregation model as main tool. 

\begin{figure*}[th]\graphicspath{{figures/rgv_vs_yuv/}}
\centering
\includegraphics[width=4.5cm]{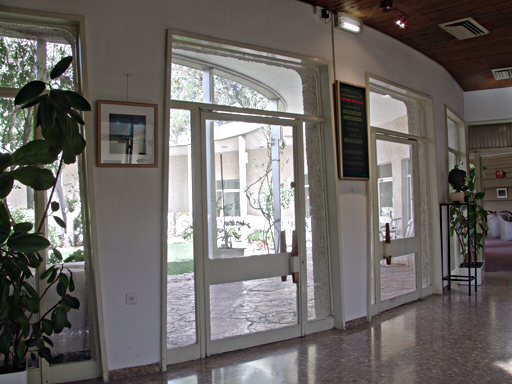} 
\includegraphics[width=4.5cm]{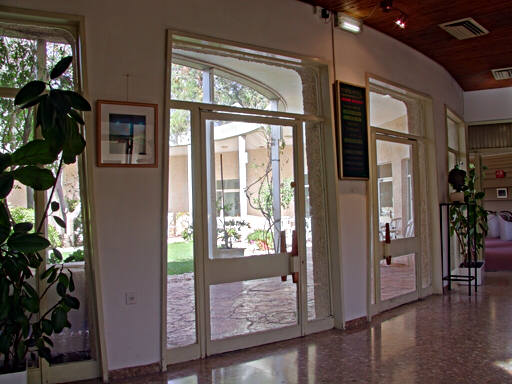} 

\caption{Comparison between applying the fusion to each RGB channel independently and using  the YUV  luminance and chromatic components. 
The luminance of the  two results is identical, while the color is enhanced by the latter strategy.  } \label{fig:rgv_vs_yuv}
\end{figure*}

\subsection{Single Channel image fusion}

Let assume we have a multi-exposure sequence of image luminance, supposed to be pre-registered, which we denote by $Y_k$, $k=1, 2,\dots, K$.
Since each image might contain well exposed areas, we apply the combination locally.
We split the images  $Y_k$ into partially-overlapped  blocks of $b\times b$ pixels, $\{B_k^l\}, \ l=1,\dots, n_b$, $n_b$ the number of blocks. 

{Both} under { and} over exposed patches will have Fourier coefficients of smaller magnitude, due to the lack of high frequency information; while well exposed patches will have Fourier coefficients of larger magnitude. We propose to fuse the non-zero frequencies  of each block as follows:

\begin{equation}\label{proposed}
\hat{B}^l(\xi)=\sum_{k=1}^K w_k^l(\xi) \hat{B}_k^l(\xi), \quad \xi\not =0, \quad l=1,2,\dots, n_b,
\end{equation}
where 
{
$\hat{B}_k^l$ denotes the DCT transform of patch $l$ in image $k$, and
}
the weights $w_k^l(\xi)$ are defined depending on $\xi$,  
\begin{equation}\label{weights}
w_k^l(\xi)=\frac{|\hat{B}_k^l(\xi)|^p}{\sum_{n=1}^K |\hat{B}_n^l(\xi)|^p}  \qquad \xi \not=0 \, .
\end{equation}
The parameter $p$ controls the weight of each Fourier mode. 

This strategy does not apply to the zero frequency Fourier coefficient, $\xi=0$, i.e. the mean. Since large zero frequency coefficients correspond to over-exposed images,  applying the same weighted combination would simply overexpose the fused image.  
{We weight these coefficients depending on how exposed is the local patch, as well as the entire image{, with} respect to the other ones. 
\begin{equation*}
\hat{B}^l(0)=\sum_{k=1}^K w_k^l(0) \hat{B}_k^l(0), \quad l=1,2,\dots, n_b,
\end{equation*}
where 
\begin{equation}\label{eq:exposure}
w_k^l(0)=\frac{1}{C} e^{-( \hat{B}_k^l(0) - 0.5)^2 / \sigma_l^2} \cdot  e^{-( \mu_k - 0.5)^2 / \sigma_g^2}  ,
\end{equation}
with $\mu_k$ the  average of  image $k$, $C$ a normalizing constant{, and $\sigma_l$ and
$\sigma_g$ parameters of the method.} The averages of the patches are normalized to $[0,1]$, then $0.5$ corresponds to the gray level mid value.

Finally, applying the inverse DCT transform to each block we obtain the fused blocks.
\begin{equation}
B^l(x)={\mathcal{F}}^{-1}(\hat{B^l}(\xi)), \quad l=1,\dots, n_b.
\end{equation}
  
Since blocks are partially overlapped, the pixels in overlapping areas are averaged to produce the final image.}

\subsection{Color image fusion}

We use an orthonormal chromatic version of the well known YUV space, described by the following linear transformation. 
$$\left(\begin{array}{c}
\mbox{Y} \\ \mbox{U} \\ \mbox{V}
\end{array} \right) = \left( \begin{array}{ccc} \frac{1}{\sqrt{3}} & \frac{1}{\sqrt{3}}&  \frac{1}{\sqrt{3}}\\ \frac{\sqrt{2}}{2} & 0 &  -\frac{\sqrt{2}}{2} \\ \frac{1}{\sqrt{6}} & -\frac{2}{\sqrt{6}} & \frac{1}{\sqrt{6}}  \end{array} \right)\left(\begin{array}{c}
\mbox{R} \\ \mbox{G} \\ \mbox{B}
\end{array} \right)$$
The Y channel, given by the average of the RGB values, represents the luminance, while U and V contain the chromatic information.  The use of an orthonormal transform is motivated by the denoising stage described below. 

The channel Y is processed as exposed above, with the average coefficient combined depending on exposure.
However, for the U and V components, we use the 
{ weighted average defined by Equations~\eqref{proposed} and~\eqref{weights}}
for all coefficients, including $\xi=0$.

The use of such a weighting does not increase the risk of over-exposure, but enhances the patch average chromaticity, making the result more colorful than { when} applying the single channel method to each of the components R, G and B. This is noticeable in Fig. \ref{fig:rgv_vs_yuv}, in which both strategies are compared.

\subsection{Noise removal}

Assuming a classical Gaussian uniform white noise scenario with standard deviation $\sigma$, the proposed fusion method can naturally incorporate noise removal. By modifying the weight definition $\eqref{weights}$ to
\begin{equation}\label{weightsNoise}
w_k^l(\xi)=\frac{Thr_\sigma(|\hat{B}_k^l(\xi)|)^p}{\sum_{n=1}^K Thr_\sigma(|\hat{B}_n^l(\xi)|)^p},  \qquad \xi \not=0
\end{equation}
where
{
$$Thr_\sigma(|\hat{B}_k^l (\xi)|) = \left\{ \begin{array}{cc}  0 & |\hat{B}_k^l (\xi)| < {T}\cdot \sigma \\ & \\ |\hat{B}_k^l (\xi)| & \mbox{otherwise} \end{array} \right.$$
}

Since the YUV color transformation is orthonormal, { the} noise standard deviation is not modified by the linear transformation converting from RGB to the mentioned space. { Thus, the same thresholding can be applied to each channel.} 
{ The parameter {$T$} is set to 2.7 as usual when denoising by thresholding in an orthonormal basis.}

\begin{figure*}  \graphicspath{{figures/example_office_s15/}}
\centering
\begin{tabular}{@{}c@{\hskip 0.4em}c@{\hskip 0.4em}c@{}}
\includegraphics[width=3.5cm]{init.i1.png}  &
\includegraphics[width=3.5cm]{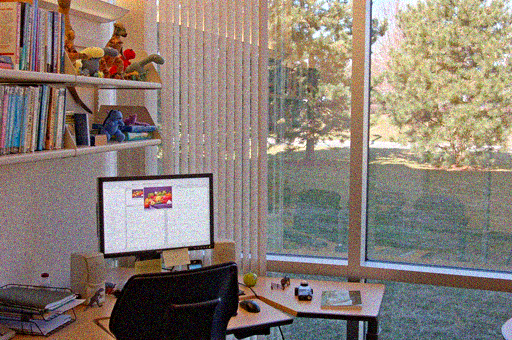} & 
\includegraphics[width=3.5cm]{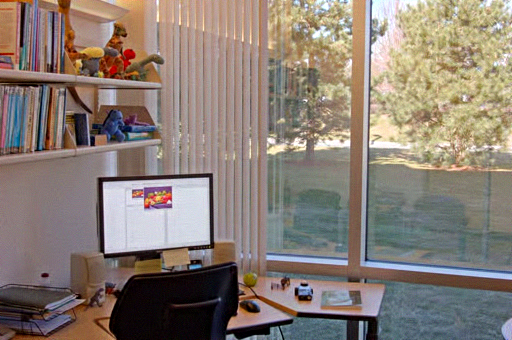} \\
\includegraphics[trim=200 100 70 30, clip,width=3.5cm]{init.i1.png}  &
\includegraphics[trim=200 100 70 30, clip,width=3.5cm]{out.mefInit.c1.t0.y.png} & 
\includegraphics[trim=200 100 70 30, clip,width=3.5cm]{out.mefInit.c1.t40.y.png} \\
\scriptsize One of the exposures & \scriptsize Fusion & \scriptsize Fusion with DCT thresh. \\
\end{tabular} 

\caption{Noisy example with sequence Office and noise { standard deviation} $15$. Top and from left to right: one image from the exposures set, fused without and with DCT thresholding. Below: detail of the images. }\label{fig:fusion_with_thr}
\end{figure*}

Fig. \ref{fig:fusion_with_thr} compares the application of the fusion chain with and without this DCT thresholding stage.

{ 
When dealing with high levels of noise,
the DCT thresholding method described above is not enough to provide good denoising results.
The next section describes how the fusion method can be combined with a collaborative
denoising technique to obtain much better results.}

%

\section{Joint noise removal and fusion procedure}\label{sec:bm3d}

\begin{figure*}[th!]
\centering
\includegraphics[width=12cm]{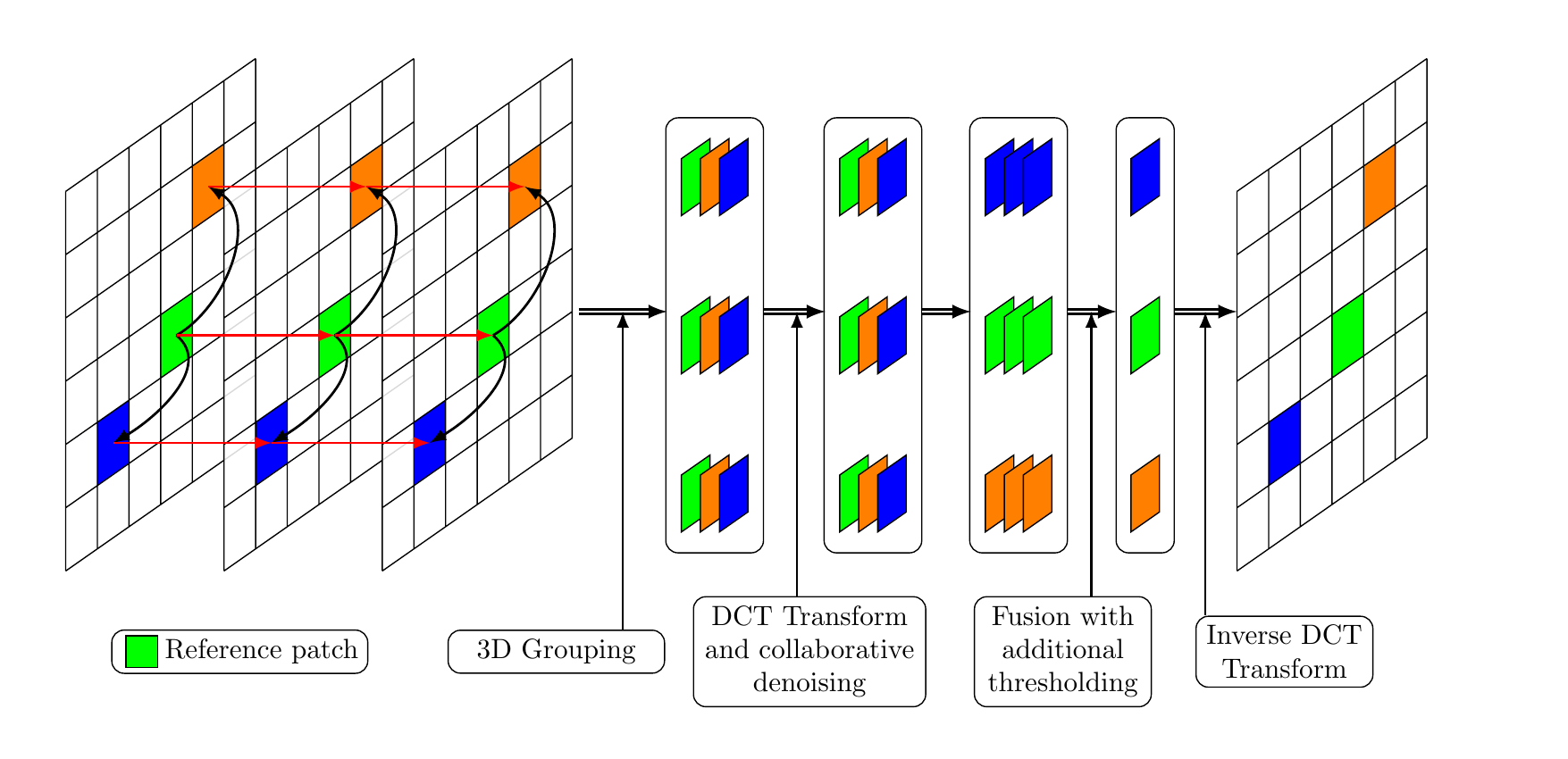}

\caption{Processing scheme of a specific reference patch. }\label{fig:scheme}
\end{figure*}

We assume as in the previous sections, that data is static and a uniform white noise model of 
 { standard deviation} $\sigma$.
Many variants have been proposed to deal with the noise removal of image sequences having the same exposure and noise conditions. 
Such methods cannot be directly used for MEF.

We propose a joint noise removal and fusion procedure (see scheme in Fig. \ref{fig:scheme}). The noise removal stage uses a collaborative   { DCT thresholding} as in \cite{dabov2007image}, and a spatio-temporal patch selection as in \cite{buades2016patch}.
The spatio-temporal selection { permits} a more robust patch comparison.  
The use of collaborative DCT thresholding permits a natural integration with the proposed fusion method, 
{ since both denoising and fusion are performed in the DCT domain,
thus avoiding the need to reconstruct each denoised image prior to the fusion step.
}

The proposed strategy   { is well adapted to} the MEF framework, { namely to the fact that all the images have different exposure}. { In the patch-selection process,} only patches belonging to the same image are compared, thus avoiding computing the similarity of patches with different exposure.  In the same line, only patches having the same exposure are  { used in the denoising step.}

{ 
Given the static image sequence, we build 3D blocks of patches, composed by the 
2D patches from the sequence located at the same spatial position.
The distance between different 3D blocks is computed as
\[
d_\text{3D}(\bm{x}, \bm{y}) = \sum_{\text{image $i$ in sequence}} ||P_i(\bm{x})-P_i(\bm{y})||
\]
\noindent
where $P_i(\bm{x})$ and $P_i(\bm{y})$ denote the 2D patches referenced by $\bm{x}$ and $\bm{y}$ in image $i$ (each 2D patch is referenced by its top-left vertex). 
}

By selecting the $k$ nearest {neighboring} 3D blocks, we actually select $k$ patches for each exposure.  Collaborative DCT thresholding {(see~\cite{dabov2007image, lebrun_bm3d})} is applied separately { to} the selected 2D patches belonging to the same exposure image. 

The above collaborative filtering recovers the denoised DCT patches which are fed directly into the fusion procedure. That is, the proposed algorithm does not need to apply the 2D inverse DCT to the grouped patches  in order to recover each denoised exposure.

Each 3D block of the $k$ selected ones contains denoised 2D DCT patches corresponding at the same spatial location and different exposure.  Each one is combined by the fusion process which includes the additional DCT thresholding.
The inverse DCT is applied  { to} the {$k$} fused  patches, which permits an additional aggregation in the image domain. 
 
The joint { patch} selection procedure reduces the dependence of noise in patch comparison, improving the robustness of the method and reducing the usual artifacts of collaborative filtering.

\section{Discussion and experimental results}\label{sec:discussion}

\graphicspath{{figures_2/fusedImages/}}
\begin{figure*}[h!]
\centering
\begin{tabular}{@{}c@{\hskip 0.4em}c@{\hskip 0.4em}c@{\hskip 0.4em}c@{}}
\includegraphics[width=4.cm]{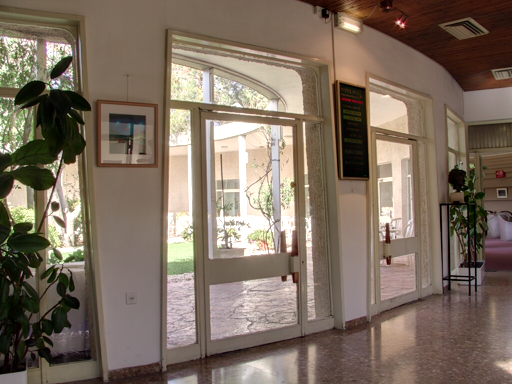} &
\includegraphics[width=4.cm]{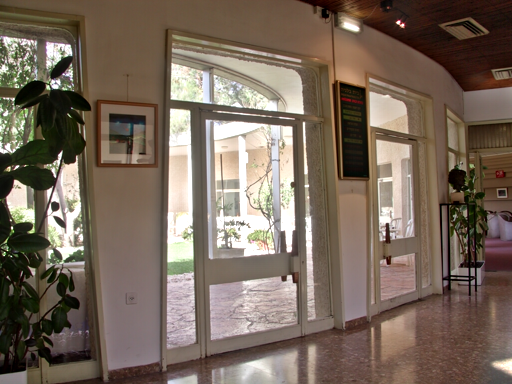} &
\includegraphics[width=4.cm]{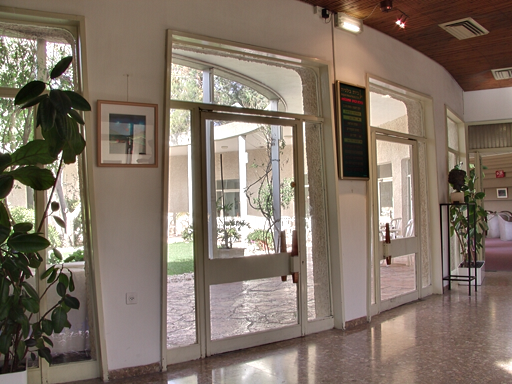} &
\includegraphics[width=4.cm]{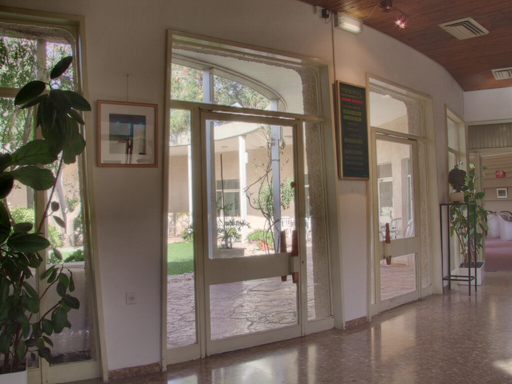} \\

Mertens et al. \cite{mertens_article}  & Ma et al. \cite{Ma17} & Li et al. \cite{Li17} & Paul et al. \cite{Paul2016MultiExposureAM} \\
\includegraphics[width=4.cm]{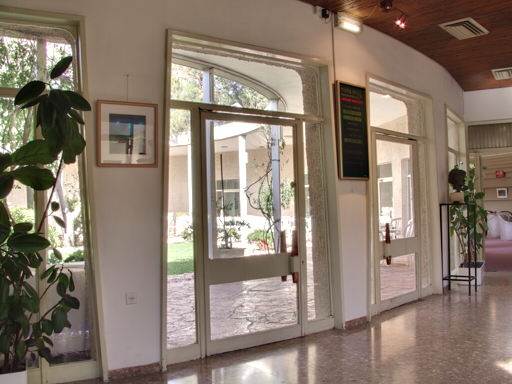} &
\includegraphics[width=4.cm]{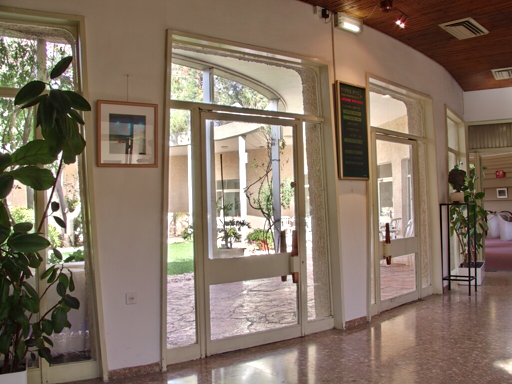} &
\includegraphics[width=4.cm]{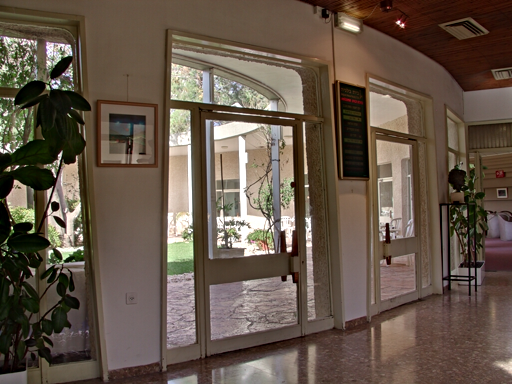} & 
\includegraphics[width=4.cm]{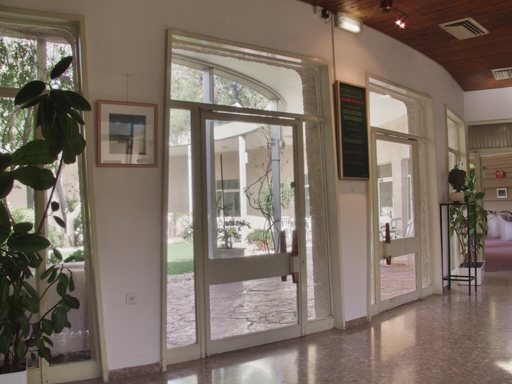} \\ 
{Kou et al. \cite{kou2017multi}} &
{Ma et al. \cite{ma2018multi} } &
Martorell et al. \cite{martorell2019ghosting} & Li et al.  \cite{8451689} \\

\includegraphics[width=4.cm]{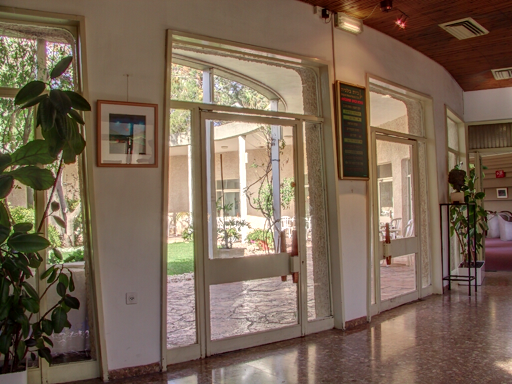} &
\includegraphics[width=4.cm]{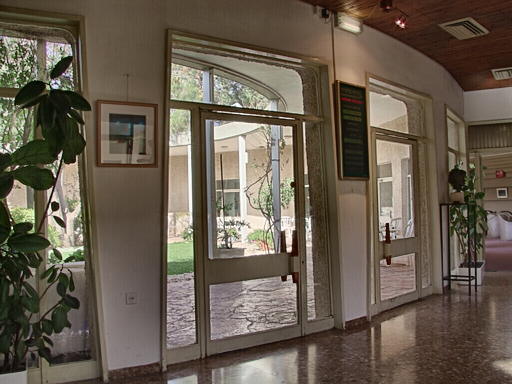} &
\includegraphics[width=4.cm]{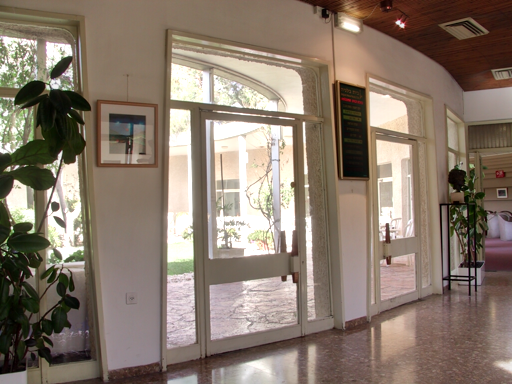} & 
\includegraphics[width=4.cm]{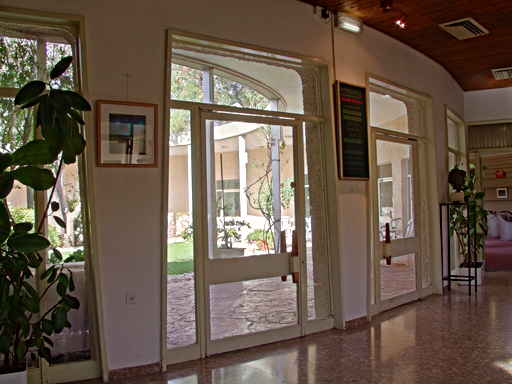} \\ 
Hessel et al \cite{ipol.2019.278} & Zhang et al. \cite{ZHANG202099} & Hayat et al. \cite{HAYAT2019295} & Ours\\
\end{tabular}

\caption{{ Results of fusion of noise-free multi-exposure images with different methods.}} \label{fig:belgium}
\end{figure*}

\graphicspath{{figures_2/fusedImages/}}
\begin{figure*}[h!]
\centering
\begin{tabular}{@{}c@{\hskip 0.4em}c@{\hskip 0.4em}c@{\hskip 0.4em}c@{}}
\includegraphics[trim=150 100 140 70, clip, width=4.0cm]{BelgiumHouse_Mertens07} &
\includegraphics[trim=150 100 140 70, clip, width=4.0cm]{BelgiumHouse_Zhang17} &
\includegraphics[trim=150 100 140 70, clip, width=4.0cm]{BelgiumHouse_DetailEnhMEF} &
\includegraphics[trim=150 100 140 70, clip, width=4.0cm]{belgium_gradientfusion} \\

Mertens et al. \cite{mertens_article}  & Ma et al. \cite{Ma17} & Li et al. \cite{Li17} & Paul et al. \cite{Paul2016MultiExposureAM} \\
\includegraphics[trim=150 100 140 70, clip, width=4.0cm]{BelgiumHouse_GGIF_MEF} &
\includegraphics[trim=150 100 140 70, clip, width=4.0cm]{BelgiumHouse_MaMEF_SSIM_c} &
\includegraphics[trim=150 100 140 70, clip, width=4.0cm]{BelgiumHouse_Martorell} & 
\includegraphics[trim=150 100 140 70, clip, width=4.0cm]{belgium_cnnfeat} \\ 
{Kou et al. \cite{kou2017multi}} &
{Ma et al. \cite{ma2018multi} } &
Martorell et al. \cite{martorell2019ghosting} & Li et al.  \cite{8451689} \\

\includegraphics[trim=150 100 140 70, clip, width=4.0cm]{belgium_eef} &
\includegraphics[trim=150 100 140 70, clip, width=4.0cm]{belgium_ifcnn} &
\includegraphics[trim=150 100 140 70, clip, width=4.0cm]{belgium_mefsift} & 
\includegraphics[trim=150 100 140 70, clip, width=4.0cm]{belgium_mef_new} \\ 
Hessel et al \cite{ipol.2019.278} & Zhang et al. \cite{ZHANG202099} & Hayat et al. \cite{HAYAT2019295} & Ours\\
\end{tabular}

\caption{{Extract of the results shown} 
in Fig. \ref{fig:belgium}.}\label{fig:belgium_extract}
\end{figure*}

In this section we compare the proposed method with state of the art algorithms for exposure fusion. We compare with 
 Mertens et al. \cite{mertens_article}, 
 Ma et al. \cite{Ma17}, Li et al.
 \cite{Li17}, Kou et al.
 \cite{kou2017multi}, Ma et al.
 \cite{ma2018multi}, Hessel et al, (EEF) \cite{ipol.2019.278}, Paul et al. \cite{Paul2016MultiExposureAM},  Zhang et al. (IFCNN) \cite{ZHANG202099}, Li et al. (CNNFEAT) \cite{8451689}, Hayat et al. (MEF-Sift) \cite{HAYAT2019295} and Martorell et al. \cite{martorell2019ghosting}. The results from Ma et al. \cite{Ma17} and Ma et al.
 \cite{ma2018multi} were computed with the software downloaded from the corresponding author's webpage. 
 {The results of Mertens et al. \cite{mertens_article} were obtained from the dataset provided in \cite{zeng14} and \cite{ma2015perceptual}. The code for  Hessel et al.\footnote{\url{http://www.ipol.im/pub/art/2019/278/?utm_source=doi}} (EEF) \cite{ipol.2019.278}, Paul et al.\footnote{\url{https://github.com/sujoyp/gradient-domain-imagefusion}} \cite{paul2016multi},  Zhang et al.\footnote{\url{https://github.com/uzeful/IFCNN}} (IFCNN) \cite{ZHANG202099}, Li et al.\footnote{\url{https://github.com/xiaohuiben/MEF-CNN-feature}} (CNNFEAT) \cite{8451689} and Hayat et al.\footnote{\url{https://github.com/ImranNust/Source-Code}} (MEF-Sift) \cite{HAYAT2019295} were obtained from corresponding github webpage.}
The results from Li et al. 
 \cite{Li17}, Kou et al.  \cite{kou2017multi} and Martorell et al. \cite{martorell2019ghosting} were computed with the code provided by the authors. In all cases, default parameter settings are adopted.

Our results were computed using the same parameters for all tests in this section. We use blocks of size $8\times8$ pixels and $p=7$ as the power exponent of the coefficient magnitudes {in Equation~\ref{weights}}. A sliding window approach is applied for the DCT based {denosing/}fusion. Once it is processed, the window is moved along both  directions with a displacement step of $N_\text{step}=2$.  The fact that the whole window is fused permits the processing of all the pixels in the image.

\subsection{Noise free sequences}

In this case, we just compare the ability of fusing the different exposure images.  
We use for our method just the fusion algorithm described in section {\ref{sec:proposed}}, { but}  without applying any thresholding of the DCT. { Note that none of the compared methods include this noise filtering step.} 

Fig. \ref{fig:belgium} displays the results of all the methods  { on} the ``Belgium House'' data set.  Most methods have a good global illumination. 
However, looking closer to the details in Fig. \ref{fig:belgium_extract}, we might observe that many outdoor details in Mertens et al. \cite{mertens_article} and Ma et al. \cite{Ma17} are overexposed, losing its definition.  Li et al. \cite{Li17} and Kou et al. \cite{kou2017multi} are not able to mantain the letters on the blackboard on the right side of Fig. \ref{fig:belgium_extract} and Ma et al. \cite{ma2018multi} is not able to preserve the details on the tree at the top of the detail image.  Li et al.  \cite{8451689} fusion is over-smoothed, while the result by  Hayat et al. \cite{HAYAT2019295} is over-saturated at bright parts. The result of \cite{martorell2019ghosting}, Hessel et al \cite{ipol.2019.278}, Zhang et al. \cite{ZHANG202099} and ours are quite similar.


\subsection{Noisy sequences with white uniform Gaussian Noise} 

\begin{figure*}[h!]
\centering
\begin{tabular}{@{}c@{\hskip 0.4em}c@{\hskip 0.4em}c@{\hskip 0.4em}c@{}}
\includegraphics[width=4.cm]{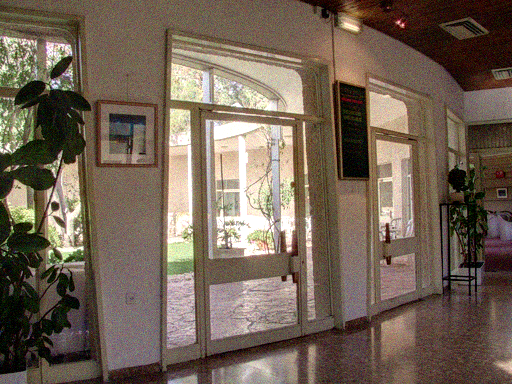} &
\includegraphics[width=4.cm]{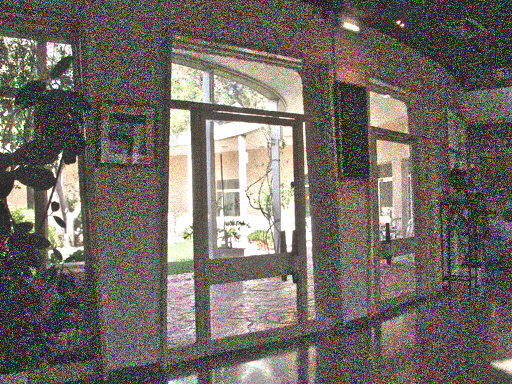} &
\includegraphics[width=4.cm]{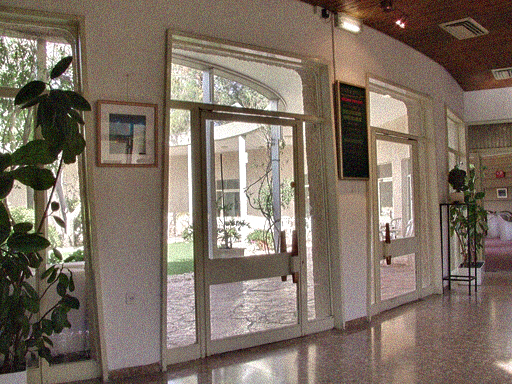} &
\includegraphics[width=4.cm]{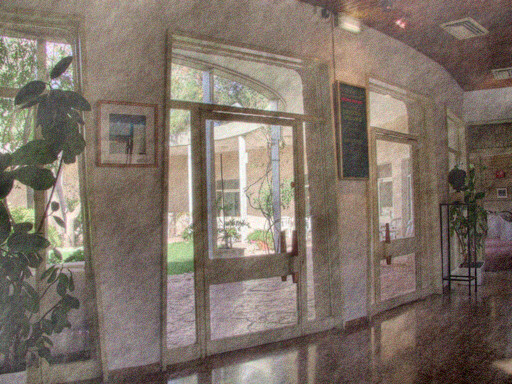} \\

Mertens et al. \cite{mertens_article}  & Ma et al. \cite{Ma17} & Li et al. \cite{Li17} & Paul et al. \cite{Paul2016MultiExposureAM} \\
\includegraphics[width=4.cm]{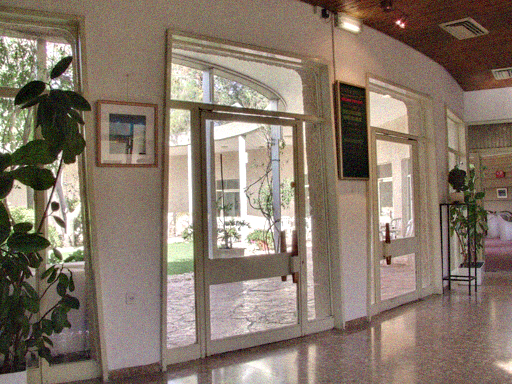} &
\includegraphics[width=4.cm]{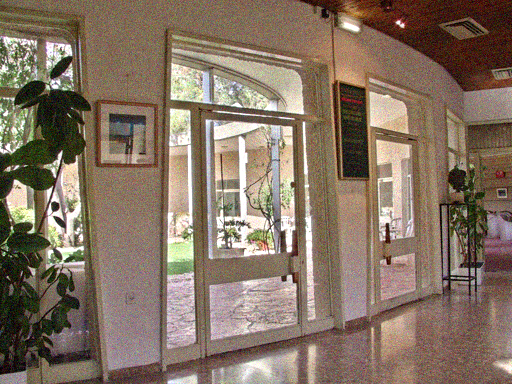} &
\includegraphics[width=4.cm]{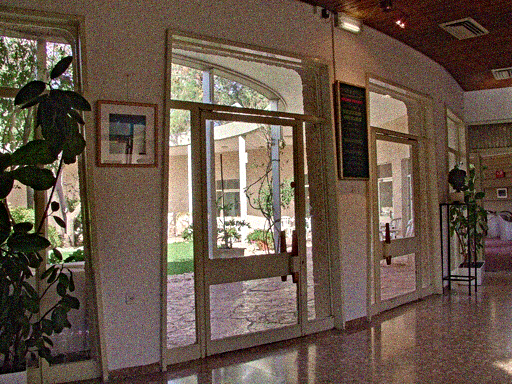}& 
\includegraphics[width=4.cm]{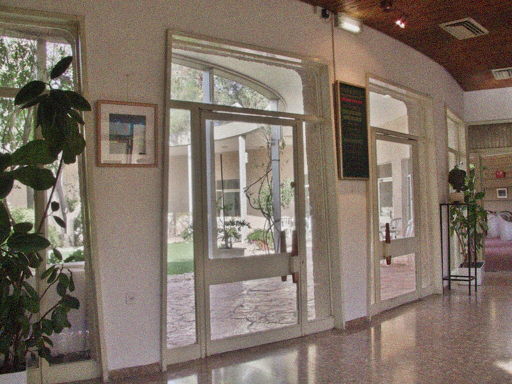} \\ 
{Kou et al. \cite{kou2017multi}} &
{Ma et al. \cite{ma2018multi} } &
Martorell et al. \cite{martorell2019ghosting} & Li et al.  \cite{8451689} \\

\includegraphics[width=4.cm]{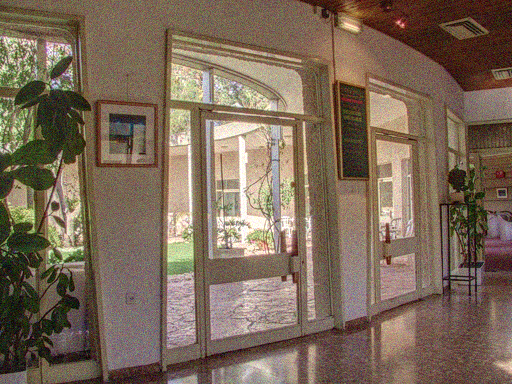} &
\includegraphics[width=4.cm]{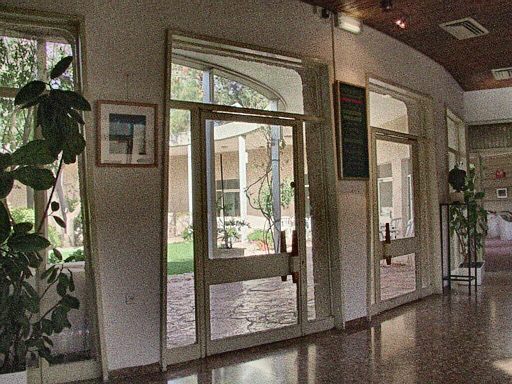} &
\includegraphics[width=4.cm]{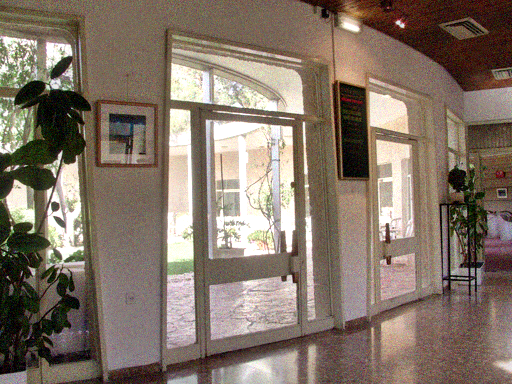} & 
\includegraphics[width=4.cm]{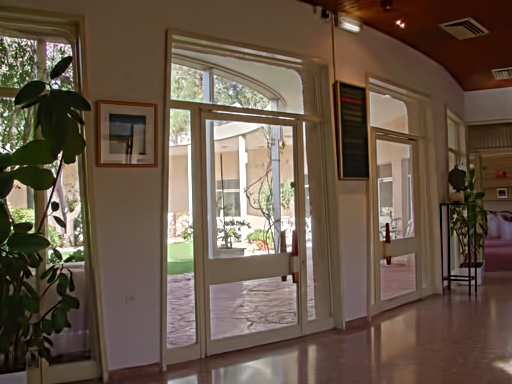} \\ 
Hessel et al \cite{ipol.2019.278} & Zhang et al. \cite{ZHANG202099} & Hayat et al. \cite{HAYAT2019295} & Ours\\
\end{tabular}

\caption{{ Results of fusion of noisy multi-exposure images with different methods. The noise standard deviation of each input image is $15$.}
} \label{fig:belgiumNoisys15}
\end{figure*}

We compare in Fig.  \ref{fig:belgiumNoisys15} all the algorithms on a noisy multi exposure sequence with standard deviation of $15$. We apply all the algorithms with their default parameters. It is clear from this figure, that the rest of methods are not adapted or do not take into account noise. 

\begin{figure*}[h!]
\centering
\begin{tabular}{@{}c@{\hskip 0.4em}c@{\hskip 0.4em}c@{\hskip 0.4em}c@{}}
\includegraphics[width=4.cm]{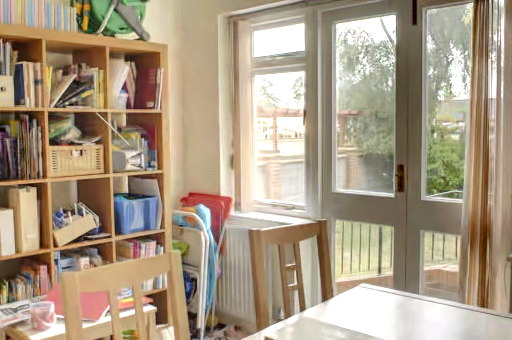} &
\includegraphics[width=4.cm]{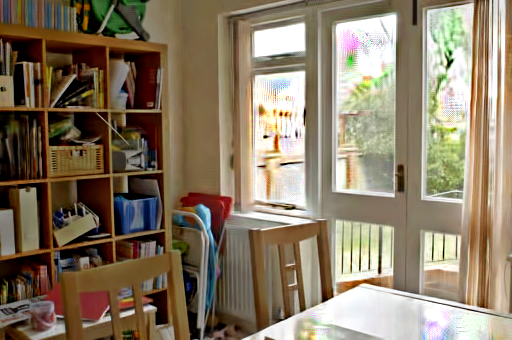} &
\includegraphics[width=4.cm]{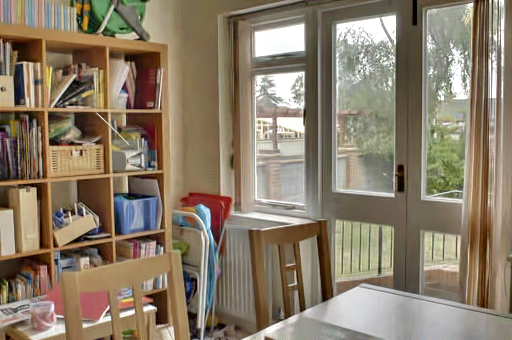} &
\includegraphics[width=4.cm]{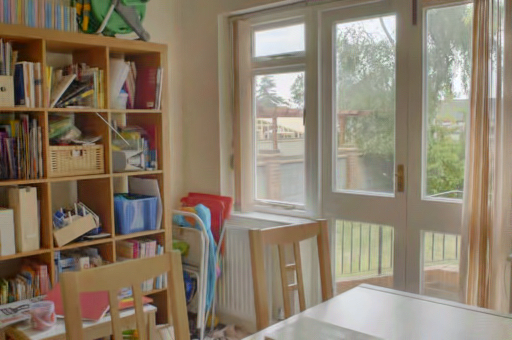} \\

Mertens et al. \cite{mertens_article}  & Ma et al. \cite{Ma17} & Li et al. \cite{Li17} & Paul et al. \cite{Paul2016MultiExposureAM} \\
\includegraphics[width=4.cm]{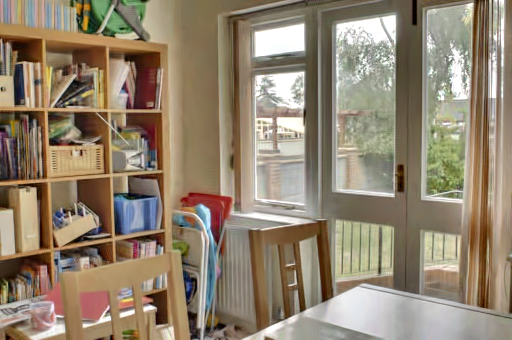} &
\includegraphics[width=4.cm]{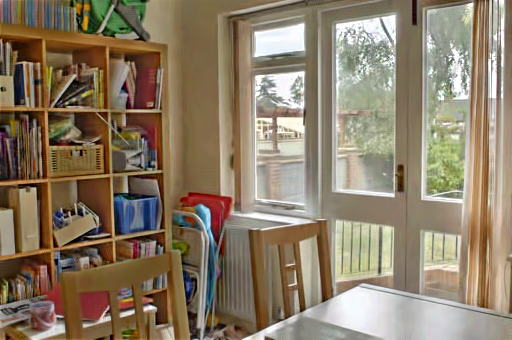} &
\includegraphics[width=4.cm]{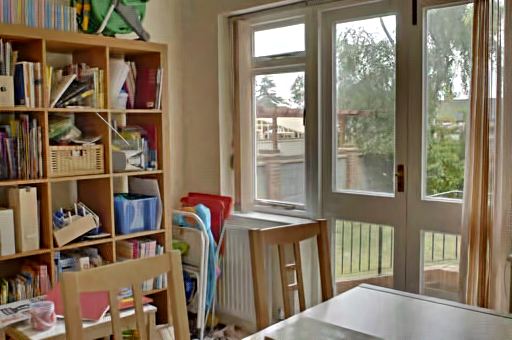}& 
\includegraphics[width=4.cm]{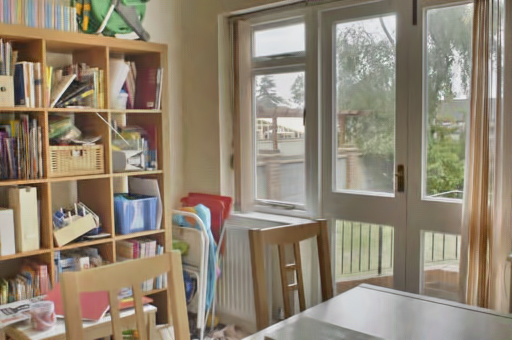} \\ 
{Kou et al. \cite{kou2017multi}} &
{Ma et al. \cite{ma2018multi} } &
Martorell et al. \cite{martorell2019ghosting} & Li et al.  \cite{8451689} \\

\includegraphics[width=4.cm]{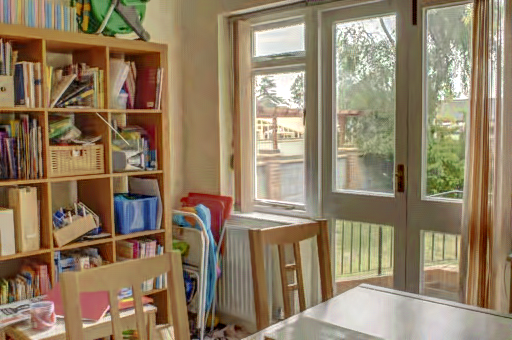} &
\includegraphics[width=4.cm]{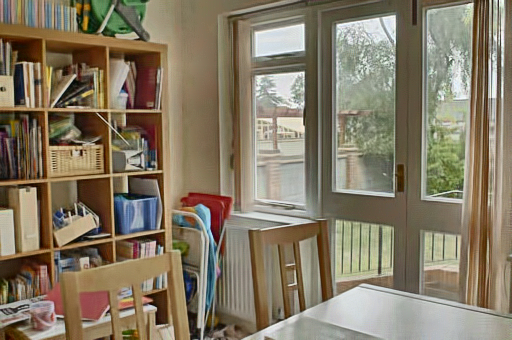} &
\includegraphics[width=4.cm]{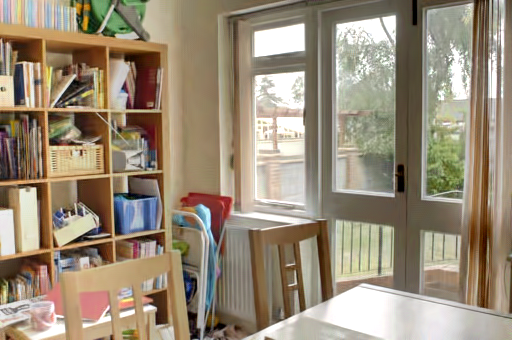} & 
\includegraphics[width=4.cm]{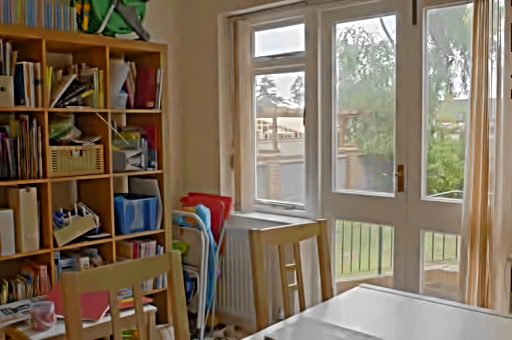} \\ 
Hessel et al \cite{ipol.2019.278} & Zhang et al. \cite{ZHANG202099} & Hayat et al. \cite{HAYAT2019295} & Ours\\
\end{tabular}

\caption{{ Results of fusion and denoising. Our method is the only one applied directly to the noisy 
multi-exposure images. The rest of methods fuse denoised versions of the images obtained using the BM3D algorithm.
The noise standard deviation of each input image is $25$.}
} \label{fig:houseNoisys25denoised}
\end{figure*}

In Fig.  \ref{fig:houseNoisys25denoised} we add noise of standard deviation 25 to each exposure and apply the  BM3D { algorithm} \cite{dabov2007image} to denoise { each image}. We apply  the { different} multi exposure fusion methods to the denoised data. Our method is applied directly to the noisy sequence.  

The zoom of an extract, Fig. \ref{fig:houseNoisys25denoisedExtract}, shows that our method is the only one able to denoise and fuse the multi exposure sequence without noticeable art{i}facts.

\subsection{Realistic multi exposure noisy images} 

We finally test our fusion and noise removal method with a realistic noise case. 
We add a signal dependent noise at each RAW image of the multi exposure set. We then apply a standard demosaicking, color processing, gamma correction and compression to the noisy raw images. The result is a realistic set of noisy multi exposure color images. 

\begin{figure*}[h!]
\centering
\includegraphics[width=4.cm]{init.i0.png} 
\includegraphics[width=4.cm]{init.i2.png} 
\includegraphics[width=4.cm]{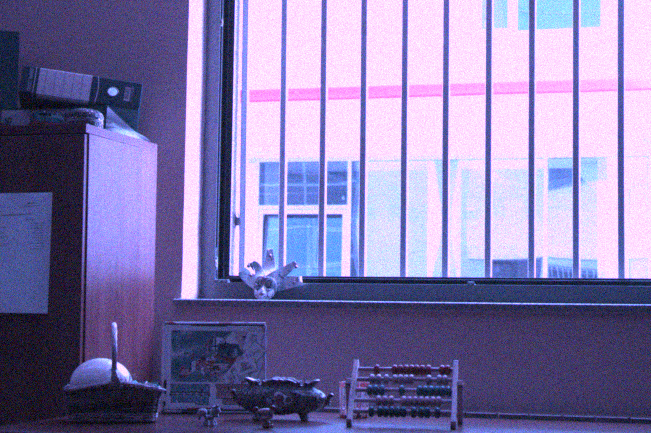} 
\includegraphics[width=4.cm]{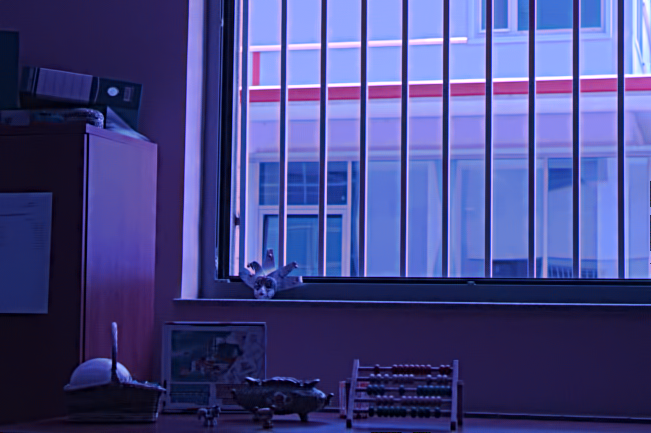}

\vspace{0.1 cm}

\includegraphics[trim=150 0 195 100, clip, width=4.cm]{init.i0.png} 
\includegraphics[trim=150 0 195 100, clip, width=4.cm]{init.i2.png} 
\includegraphics[trim=150 0 195 100, clip, width=4.cm]{init.i4.png} 
\includegraphics[trim=150 0 195 100, clip, width=4.cm]{setup4.a1.h1.v2.l0.5.f2.7.t20.s12}

\caption{Realistic noise simulation. From left to right: three exposure noisy images and the fusion and noise removal result. Below an extract of each image.} \label{fig:realistic}
\end{figure*}

We apply our denoising and fusion algorithm to  { these} images. The result is displayed in Fig. \ref{fig:realistic}.

\subsection{Computational analysis}

{

The time complexity of the proposed algorithm is $\mathcal{O}(|Y|)$, 
where $|Y|$ denotes the size of each image in the multi-exposure sequence.

Assuming that the 3D transforms used for the collaborative filtering are performed
in a separable way (i.e. 2D transforms followed by 1D transforms, as detailed in~\cite{dabov2007image}), 
the overall number of operations of the algorithm, per pixel, is approximately 

\begin{equation}
{\cal C}_{{\cal T}_{2D}} + 2Kb^2N_S^2 + 2Kb^2 {\cal C}_{{\cal T}_{1D}}  +  2Kkb^2 + k {\cal C}'_{{\cal T}_{2D}} + k b^2
\label{eq:complexity}
\end{equation}
\noindent
where:
\begin{itemize}
\item ${\cal C}_{{\cal T}_{2D}}$ denotes the number of operations required to compute the 2D DCT
of the block of patches similar to the one centered at the considered pixel. If we consider a 
neighborhood of size $N_S \times N_S$ around the pixel, this implies the computation of $K N_S^2$ 2D DCTs,
where $K$ is the number of images in the sequence.
The time complexity can be reduced by pre-computing the transforms in each block of size $K \times N_S \times N_S$
and reusing them in overlapping blocks, similarly to what is proposed in~\cite{dabov2007image}.

\item The second term accounts for the 3D block matching step. This implies the exhaustive search, in a $N_S \times N_S$ neighborhood of the pixel, of 3D blocks of size $K \times b\times b$.

\item The third term counts the number of operations for the computation of the 1D DCT transforms (direct and inverse) of the $k$ nearest neighbors of each patch, in each frame. ${\cal C}_{{\cal T}_{1D}}$ denotes the cost of computing a 1D DCT transform (direct or inverse) of a vector of size $k$.

\item The fourth term accounts for the fusion step, which involves $k$ 3D blocks of patches of size $K \times b\times b$. 

\item The fifth term accounts for the number of operations needed to compute the $k$ inverse 2D DCT transforms of the fused patches. ${\cal C}'_{{\cal T}_{2D}}$ denotes the cost of computing the inverse 2D DCT transform of a patch of $b \times b$ pixels.

\item Finally, the last term counts the number of operations involved in the aggregation step.

\end{itemize}

Observe that the number of denoising operations per pixel, for each image, is smaller than that of BM3D since only one step of the collaborative filtering is applied. In addition, the inverse DCT is applied only to fused patches, since denoised exposures are not required by the algorithm. 

Moreover, the previous estimation assumes that an exhaustive-search algorithm has been used for block matching. 
The costs ${\cal C}_{\cal T}$ of the DCT transforms depends on the availability of fast algorithms.
By using predictive search techniques and fast separable transforms the complexity of the
algorithm could be significantly reduced. Moreover,
the overall number of operations can be further reduced by processing only one out of each $N_\text{step} < b$ pixels in both the horizontal and vertical directions. Due to the overlapping of the patches, the aggregation step used in the final step of the algorithm guarantees that all the pixels are correctly processed. In this case, the overall complexity of the method is reduced by a factor $N_\text{step}^2$.

}

\medskip

\medskip

\graphicspath{{figures/}}
\begin{figure*}[t!]
\centering

\includegraphics[width=3cm]{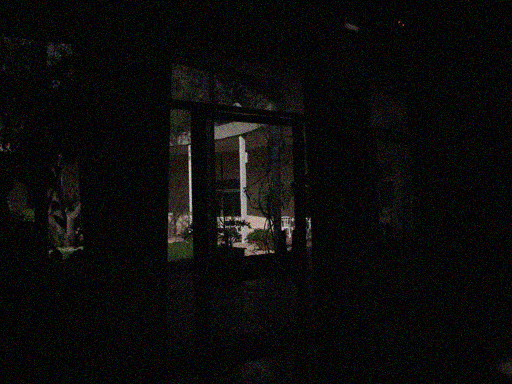} 
\includegraphics[width=3cm]{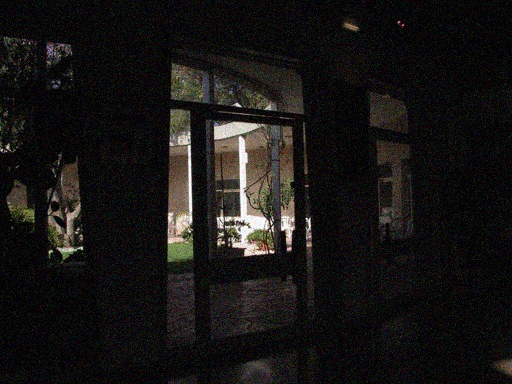} 
\includegraphics[width=3cm]{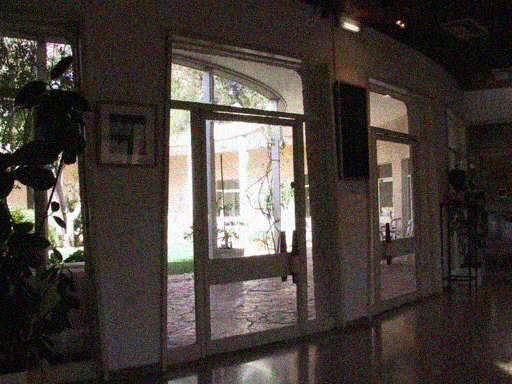} 
\includegraphics[width=3cm]{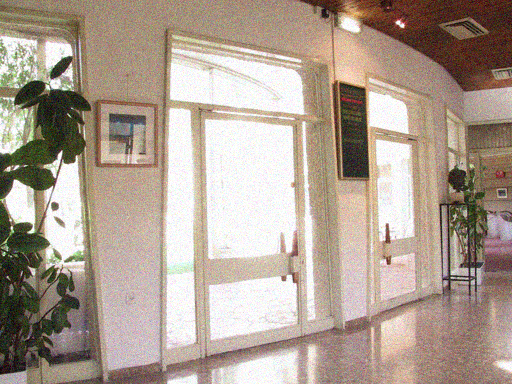} 

\includegraphics[width=3cm]{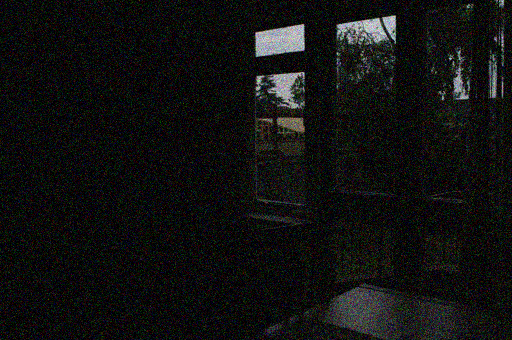} 
\includegraphics[width=3cm]{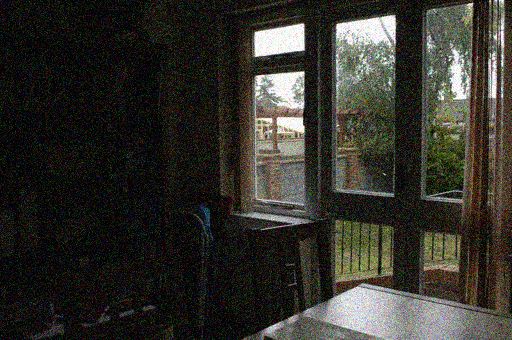} 
\includegraphics[width=3cm]{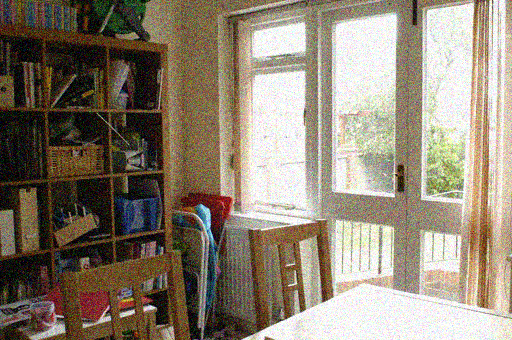} 
\includegraphics[width=3cm]{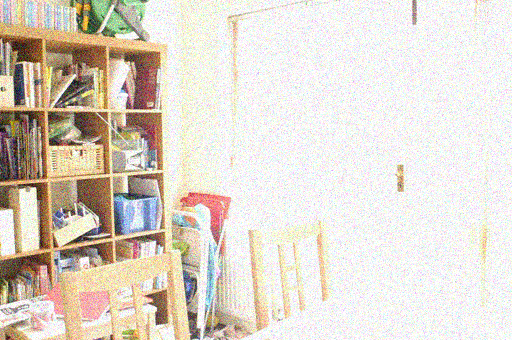}

\caption{Noisy multi-exposure data sets used for comparison. On the first row, the noise standard deviation of each input image is $15$. On the second row the standard deviation is $25$.} 
\end{figure*}

\graphicspath{{figures_2/noise/results_noisy/}}

\graphicspath{{figures_2/noise/results_noisy/}}
\begin{figure*}[h!]
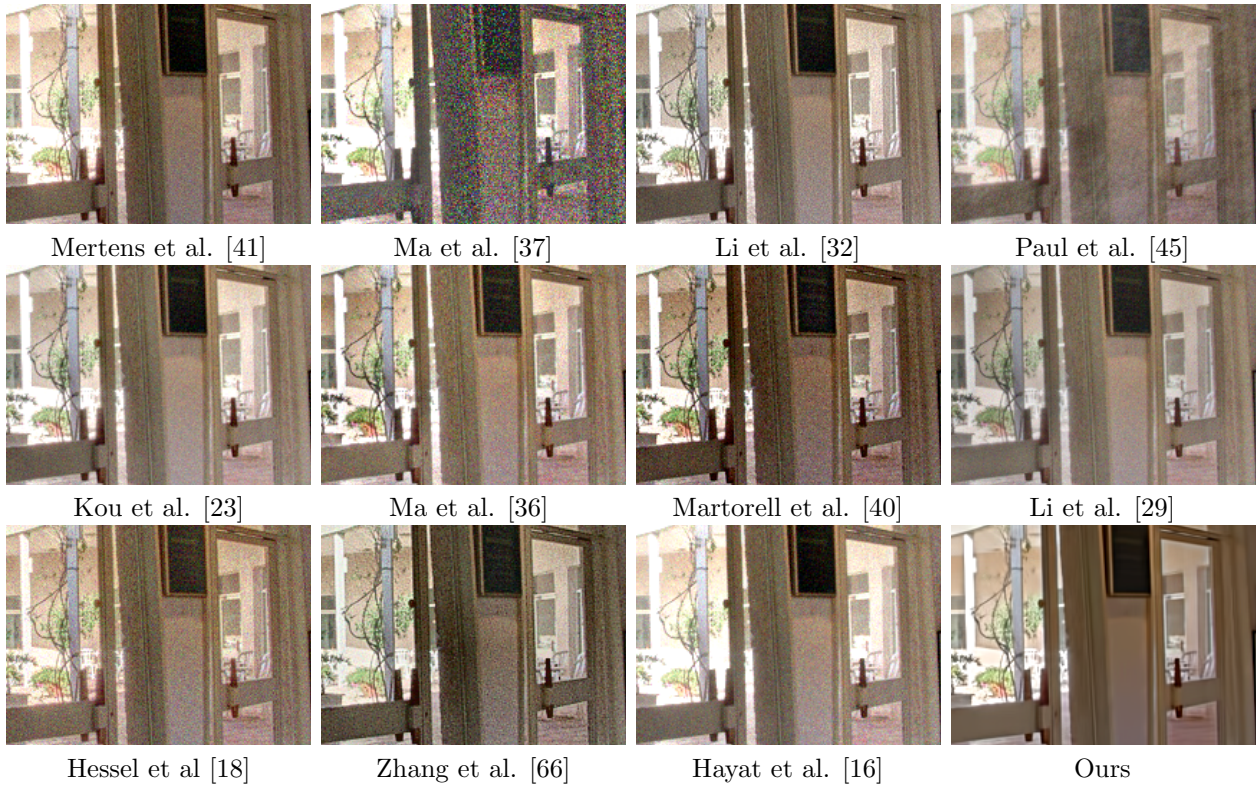

\centering
\begin{tabular}{@{}c@{\hskip 0.4em}c@{\hskip 0.4em}c@{\hskip 0.4em}c@{}}
\includegraphics[trim=230 100 70 130, clip, width=4.cm]{belgium_s15_noisy_mertens_static} &
\includegraphics[trim=230 100 70 130, clip, width=4.cm]{belgium_s15_noisy_ZhangMEF} &
\includegraphics[trim=230 100 70 130, clip, width=4.cm]{belgium_s15_noisy_DetailEnhMEF} &
\includegraphics[trim=230 100 70 130, clip, width=4.cm]{belgium_s15_noisy_gradientfusion} \\

Mertens et al. \cite{mertens_article}  & Ma et al. \cite{Ma17} & Li et al. \cite{Li17} & Paul et al. \cite{Paul2016MultiExposureAM} \\
\includegraphics[trim=230 100 70 130, clip, width=4.cm]{belgium_s15_noisy_GGIF_MEF} &
\includegraphics[trim=230 100 70 130, clip, width=4.cm]{belgium_s15_noisy_MaMEF_SSIM_c} &
\includegraphics[trim=230 100 70 130, clip, width=4.cm]{belgium_s15.pubUsingMertens.r9}& 
\includegraphics[trim=230 100 70 130, clip, width=4.cm]{belgium_s15_noisy_cnnfeat} \\ 
{Kou et al. \cite{kou2017multi}} &
{Ma et al. \cite{ma2018multi} } &
Martorell et al. \cite{martorell2019ghosting} & Li et al.  \cite{8451689} \\

\includegraphics[trim=230 100 70 130, clip, width=4.cm]{belgium_s15_noisy_eef} &
\includegraphics[trim=230 100 70 130, clip, width=4.cm]{belgium_s15_noisy_ifcnn} &
\includegraphics[trim=230 100 70 130, clip, width=4.cm]{belgium_s15_noisy_mefsift} & 
\includegraphics[trim=230 100 70 130, clip, width=4.cm]{belgium_s15.a1.h1.v2.l0.5.f2.7.t30} \\ 
Hessel et al \cite{ipol.2019.278} & Zhang et al. \cite{ZHANG202099} & Hayat et al. \cite{HAYAT2019295} & Ours\\
\end{tabular}

\caption{{ Extract of the results shown in Fig. \ref{fig:belgiumNoisys15}. It is clear from this figure, that our method is the only one that takes noise into account}
} \label{fig:belgiumNoisys15_extract}
\end{figure*}

\graphicspath{{figures_2/noise/results_denoised/}}

\graphicspath{{figures_2/noise/results_denoised/}}
\begin{figure*}[h!]
\centering
\begin{tabular}{@{}c@{\hskip 0.4em}c@{\hskip 0.4em}c@{\hskip 0.4em}c@{}}
\includegraphics[trim=200 100 70 30, clip, width=4.cm]{house_s25_bm3d_mertens_static} &
\includegraphics[trim=200 100 70 30, clip, width=4.cm]{house_s25_bm3d_ZhangMEF} &
\includegraphics[trim=200 100 70 30, clip, width=4.cm]{house_s25_bm3d_DetailEnhMEF} &
\includegraphics[trim=200 100 70 30, clip, width=4.cm]{house_s25_denoised_gradientfusion} \\

Mertens et al. \cite{mertens_article}  & Ma et al. \cite{Ma17} & Li et al. \cite{Li17} & Paul et al. \cite{Paul2016MultiExposureAM} \\
\includegraphics[trim=200 100 70 30, clip, width=4.cm]{house_s25_bm3d_GGIF_MEF} &
\includegraphics[trim=200 100 70 30, clip, width=4.cm]{house_s25_bm3d_MaMEF_SSIM_c} &
\includegraphics[trim=200 100 70 30, clip, width=4.cm]{house_s25_bm3d.pubUsingMertens.r9}& 
\includegraphics[trim=200 100 70 30, clip, width=4.cm]{house_s25_denoised_cnnfeat} \\ 
{Kou et al. \cite{kou2017multi}} &
{Ma et al. \cite{ma2018multi} } &
Martorell et al. \cite{martorell2019ghosting} & Li et al.  \cite{8451689} \\

\includegraphics[trim=200 100 70 30, clip, width=4.cm]{house_s25_denoised_eef} &
\includegraphics[trim=200 100 70 30, clip, width=4.cm]{house_s25_denoised_ifcnn} &
\includegraphics[trim=200 100 70 30, clip, width=4.cm]{house_s25_denoised_mefsift} & 
\includegraphics[trim=200 100 70 30, clip, width=4.cm]{house_s25.a1.h1.v2.l0.5.f2.7.t65} \\ 
Hessel et al \cite{ipol.2019.278} & Zhang et al. \cite{ZHANG202099} & Hayat et al. \cite{HAYAT2019295} & Ours\\
\end{tabular}

\caption{{ Extract of the results shown in Fig.~\ref{fig:houseNoisys25denoised}.} 
} \label{fig:houseNoisys25denoisedExtract}
\end{figure*}

\graphicspath{{figures_2/realcase/}}

\section{Conclusions}
\label{sec:conclusions}
{
{ In this paper we propose a patch-based method for the simultaneous 
denoising and fusion of a sequence of multi-exposure images. Both tasks are performed in the DCT domain
and take advantage of a collaborative 3D thesholding approach similar to BM3D~\cite{dabov2007image} for denoising, 
and of the state-of-the-art fusion technique proposed in~\cite{martorell2019ghosting}.
For the collaborative denoising, a spatio-temporal criterion is used to select similar patches along the sequence,
following the approach in~\cite{buades2016patch}.}
The overall strategy permits to denoise and fuse the set of images without the need of recovering each denoised exposure image, leading to a very efficient procedure. 

{ Several experiments show that the proposed method permits to obtain state-of-the-art fusion results even when the input multi-exposure images are noisy.}
As future work, we plan to extend the current approach to multi exposure video sequences.
}

\section*{Acknowledgments}
The authors thank  Zhengguo Li and Fei Kou for kindly providing the implementation of \cite{Li17} and \cite{kou2017multi}, respectively.

\bibliographystyle{plain}
\bibliography{refs}

\end{document}